\documentclass[preprint]{ptephy_v1}%%%%%% to generate preprint number

\preprintnumber{XXXX-XXXX} %%% %%% Insert preprint number here

\usepackage{hyperref}% for linking the cross references
\usepackage{lastpage}
\usepackage{booktabs}
\usepackage{multirow}
\usepackage{lineno}
\newcommand{\Var}{\mathrm{Var}}

\newcommand{\Cov}{\mathrm{Cov}}

\newcommand{\ellmax}{\ell_\mathrm{max}}

\nolinenumbers

\begin{document}
\title{Simultaneous determination of the cosmic birefringence and miscalibrated polarization angles II: Including cross frequency spectra}

%%%% To generate auto affiliation numbers please use \author{}\affil{} command

\author{Yuto Minami}
\affil{High Energy Accelerator Research Organization, 1-1 Oho, Tsukuba, Ibaraki 305-0801, Japan \email{yminami@post.kek.jp}}

\author[2,3]{Eiichiro Komatsu}
\affil[2]{Max Planck Institute for Astrophysics, Karl-Schwarzschild-Str. 1, D-85748 Garching, Germany}
\affil[3]{Kavli Institute for the Physics and Mathematics of the Universe (Kavli IPMU, WPI), Todai Institutes for Advanced Study, The University of Tokyo, Kashiwa 277-8583, Japan}

%%% To include the collaborator name... Please use the command "\collaborator"
%%% For example: \collaborator{ATLAS Collaboration}

\begin{abstract}%
We develop a strategy to determine the cosmic birefringence and miscalibrated polarization angles simultaneously using the observed $EB$ polarization power spectra of the cosmic microwave background and the Galactic foreground emission. We extend the methodology of Y. Minami et al. (Prog. Theor. Exp. Phys. {\bf 2019}, 083E02, 2019), which was developed for auto frequency power spectra, by including cross frequency spectra. By fitting one global birefringence angle and independent miscalibration angles at different frequency bands, we determine both angles with significantly smaller uncertainties (by more than a factor of two) compared to the auto spectra.
\end{abstract}
\subjectindex{xxxx, xxx}

\maketitle

\section{Introduction}\label{sec:Introduction}
Measuring the rotation of polarization directions of photons is a way to search for parity violating physics. One of the physical effects is the so-called ``cosmic birefringence''~\cite{Carroll:1998zi}, which rotates the polarization angle of the cosmic microwave background (CMB) via, e.g., a Chern-Simons coupling between a light scalar field and the electromagnetic tensor. The rotation mixes $E$- and $B$-modes of CMB polarization and generates a cross-correlation between them; thus, we can search for this effect in the cross-correlation power spectrum of the $E$- and $B$-mode polarization \cite{,Lue:1998mq,Feng:2004mq,Feng:2006dp,Liu:2006uh,Saito:2007kt}.

The polarization angle of the CMB is rotated also by miscalibration of polarization sensitive detectors.
Assuming no intrinsic $EB$ correlation, one can determine the miscalibration angles \cite{Keating:2012ge}; however, the miscalibration angles determined in this way are degenerate with the cosmic birefringence~\cite{Komatsu:2010fb}. In Ref.~\cite{Minami:2019ruj}, we lifted the degeneracy by the polarized Galactic foreground emission. The miscalibration rotates both CMB and foregrounds by an angle $\alpha$, whereas the cosmic birefringence rotates only CMB by an angle $\beta$; thus, we can use the different multipole dependence of the CMB and the Galactic foreground polarization power spectra to determine $\alpha$ and $\beta$ simultaneously.

In this paper, we extend the methodology of Ref.~\cite{Minami:2019ruj}, which was limited to the auto frequency power spectra, to include the cross frequency spectra. Modern CMB experiments have many frequency bands required to remove the foreground emission. For $N_\nu$ frequency bands, we can measure $N_\nu$ auto frequency spectra and $N_\nu(N_\nu-1)/2$ cross spectra. When $N_\nu$ is large, the information gained from the cross power spectra is significant. In Ref.~\cite{Minami:2019ruj}, we showed that an experiment similar to LiteBIRD~\cite{Hazumi2019} can determine $\beta$ with a precision of 11\,arcmin using the auto frequency power spectra, simultaneously with $\alpha$.\footnote{The formal statistical uncertainty of $\beta$ in the absence of $\alpha$ can be much smaller \cite{PhysRevD.80.043522,Gubitosi:2014cua,MOLINARI201665,Gruppuso_2016,Gruppuso:2016nhj,Pogosian:2019jbt}; however, this is not achievable in practice unless we can determine $\alpha$ with better precision than we report in this paper.}
In this paper we show that we can reduce the uncertainty by more than a factor of two by including the cross frequency spectra.

Throughout this paper we assume the full sky coverage. See Ref.~\cite{Minami:2020xfg} for how to include the effect of a partial sky coverage. 

The rest of the paper is organized as follows.
In Sect.~\ref{sec:Methodology} we review the methodology of Ref.~\cite{Minami:2019ruj} and describe the extension to the cross frequency power spectra. 
In Sect.~\ref{sec:SkySim} we describe the sky simulations we use to validate our method. 
We present the main results in Sect.~\ref{sec:Results}  and
conclude in Sect.~\ref{sec:Conclusion}.

\section{Methodology}\label{sec:Methodology}
\subsection{\label{sec:review}Review of the methodology with auto frequency power spectra}
The spherical harmonics coefficients of the observed (``o'') $E$- and $B$- mode polarization are related to the intrinsic ones by 
\begin{align}
E_{\ell,m}^\mathrm{o} &= 
E_{\ell,m}^\mathrm{fg}\cos(2\alpha)-  B_{\ell,m}^\mathrm{fg}\sin(2\alpha)
+E_{\ell,m}^\mathrm{CMB}\cos(2\alpha+2\beta)-  B_{\ell,m}^\mathrm{CMB}\sin(2\alpha+2\beta)
+E_{\ell,m}^\mathrm{N}\,,
\label{eq:Emode}
\\
B_{\ell,m}^\mathrm{o} &= 
E_{\ell,m}^\mathrm{fg}\sin(2\alpha) + B_{\ell,m}^\mathrm{fg}\cos(2\alpha)
+E_{\ell,m}^\mathrm{CMB}\sin(2\alpha+2\beta) + B_{\ell,m}^\mathrm{CMB}\cos(2\alpha+2\beta)
+B_{\ell,m}^\mathrm{N}\,,
\label{eq:Bmode}
\end{align}
where ``N'', ``fg'', and ``CMB'' denote the noise, foreground, and CMB components, respectively. 
All spherical harmonics coefficients and power spectra have been multiplied by the appropriate beam smoothing functions. 

When we define the power spectra with ensemble average as $\langle C_\ell^{XY} \rangle=(2\ell+1)^{-1}\sum_{m=-\ell}^\ell \langle X_{\ell,m}Y_{\ell,m}^* \rangle$, 
we obtain
\begin{equation}\label{eq:GeneralRotationFitting}
\begin{split}
\langle C_\ell^{EB,\mathrm{o}}\rangle =&
\frac{\tan(4\alpha ) }{2}
\left(\langle C_\ell^{EE,\mathrm{o}}\rangle-\langle C_\ell^{BB,\mathrm{o}}\rangle\right)
+
\frac{\sin(4\beta)}{2\cos(4\alpha)}\left(
\langle C_\ell^{EE,\mathrm{CMB}}\rangle - \langle C_\ell^{BB,\mathrm{CMB}}\rangle
\right)\\
&+\frac1{\cos(4\alpha)}\langle C_\ell^{EB,\mathrm{fg}}\rangle
+\frac{\cos(4\beta)}{\cos(4\alpha)}\langle C_\ell^{EB,\mathrm{CMB}}\rangle\,.
\end{split}
\end{equation}
The current data show no evidence for non-zero intrinsic $EB$ correlation from the foreground emission~\cite{planckdust:2016,planckdust:2018} or from the CMB~\cite{Aghanim:2016fhp}. We can thus determine $\alpha$ and $\beta$ simultaneously, if we ignore $\langle C_\ell^{EB,\mathrm{fg}}\rangle$ and $\langle C_\ell^{EB,\mathrm{CMB}}\rangle$, or we model them. For example, if we assume that $\langle C_\ell^{EB,\mathrm{fg}}\rangle$ is proportional to 
$\sqrt{\langle C_\ell^{EE,\mathrm{fg}}\rangle\langle C_\ell^{BB,\mathrm{fg}}\rangle}$ \cite{Abitbol:2015epq}, we can model the effect of the intrinsic foreground $EB$ correlation as an additional, frequency-dependent angle, $\gamma(\nu)$ \cite{Minami:2019ruj,Minami:2020xfg}. The intrinsic $EB$ correlation in the CMB generated prior to the last scattering has the multiple dependence distinct from the second term in Eq.~(\ref{eq:GeneralRotationFitting}) (e.g., \cite{Thorne:2017jft}). These properties would allow us to distinguish between $\alpha$, $\beta$, $\gamma(\nu)$, and 
$\langle C_\ell^{EB,\mathrm{CMB}}\rangle$. In this paper we ignore the latter two effects, as our main goal is to show the improvements offered by the cross frequency spectra.

To determine $\alpha$ and $\beta$ simultaneously from a single frequency band, we use a log-likelihood function given by
\begin{equation}\label{eq:LikelihoodGeneral}
\ln \mathcal{L}
= -\frac{1}{2}\sum_{\ell=2}^{\ell_\mathrm{max}}
\frac{
	\left[
	C_\ell^{EB,\mathrm{o} } 
	- \frac{\tan(4\alpha) }{ 2 }  \left( C_\ell^{EE,\mathrm{o} } - C_\ell^{ BB,\mathrm{o} } \right) 	
	-  \frac{ \sin(4\beta) }{ 2\cos( 4\alpha ) }
	\left(
	C_{\ell}^{EE,\mathrm{CMB,th} }b_\ell^2 - C_{\ell}^{BB,\mathrm{CMB,th} }b_\ell^2 
	\right)	
	\right]^2
}{
	\Var \left( 
	C_\ell^{ EB, \mathrm{o} } - \frac{\tan(4\alpha) }{ 2 }  \left( C_\ell^{EE,\mathrm{o} } - C_\ell^{ BB,\mathrm{o} } \right) 
	\right)
}\,,
\end{equation}
where $C_{\ell}^{EE,\mathrm{CMB,th} }b_\ell^2$ and $C_{\ell}^{BB,\mathrm{CMB,th} }b_\ell^2$ are the best-fitting $\Lambda$CDM theoretical power spectra multiplied by the beam transfer functions, $b_\ell^2$.

As for the variance in the denominator,
we use an approximate variance given by
\begin{equation}\label{eq:VarAuto}
\begin{split}
&\Var \left( C_\ell^{EB, \mathrm{o}}   - ( C_\ell^{EE, \mathrm{o}} - C_\ell^{BB, \mathrm{o}})\tan(4\alpha)/2 \right) \\
&\approx
\frac{1}{2\ell+1}C_\ell^{EE,\mathrm{o}}C_\ell^{BB,\mathrm{o}}
+\frac{\tan^2(4\alpha)}{4}\frac{2}{2\ell+1}\left[
(C_\ell^{EE,\mathrm{o}})^2
+
(C_\ell^{BB,\mathrm{o}})^2
\right]
\\
&\quad
-\tan(4\alpha)\frac{2}{2\ell+1}C_\ell^{EB,\mathrm{o}}
\left( C_\ell^{EE,\mathrm{o}} - C_\ell^{BB,\mathrm{o}} 
\right)\,.
\end{split}
\end{equation}
See Appendix of Ref.~\cite{Minami:2019ruj} for derivation of this result. We maximize Eq.~(\ref{eq:LikelihoodGeneral}) with respect to $\alpha$ and $\beta$, given $C_\ell^{ EB, \mathrm{o}}$, $\left(C^{EE,\mathrm{o}}-C^{BB,\mathrm{o}}\right)$, $C_\ell^{EE,\mathrm{CMB,th}}$, and $C_\ell^{BB,\mathrm{CMB,th}}$.

\subsection{Extension to the cross frequency power spectra} \label{sec:extension}
We extend the log-likelihood given in Eq.~(\ref{eq:LikelihoodGeneral}), which is valid for one frequency band, to $N_\nu$ multi-frequency bands. 
Let us consider correlations of three pairs ($E_i E_j$, $B_iB_j$, and $E_iB_j$) 
from the $i$-th and $j$-th frequency bands.
We assume that different bands have independent miscalibration angles, $\alpha_i$.
When we take the ensemble average of the cross correlation between two frequency bands,
we obtain
\begin{align}
\begin{pmatrix}
\langle C_\ell^{E_i E_j,\mathrm{o}}\rangle\\
\langle C_\ell^{B_i B_j,\mathrm{o}}\rangle 
\end{pmatrix} &= 
\mathbf{R}(\alpha_i , \alpha_j)\begin{pmatrix}
\langle C_\ell^{E_i E_j,\mathrm{fg}}\rangle\\
\langle C_\ell^{B_i B_j,\mathrm{fg}}\rangle 
\end{pmatrix}  
+
\mathbf{R}(\alpha_i +\beta,  \alpha_j+\beta)\begin{pmatrix}
\langle C_\ell^{E_i E_j,\mathrm{CMB}}\rangle\\
\langle C_\ell^{B_i B_j,\mathrm{CMB}}\rangle 
\end{pmatrix} 
+
\delta_{i,j}
\begin{pmatrix}
\langle C_\ell^{E_i E_i, \mathrm{N} } \rangle\\
\langle C_\ell^{B_i B_i, \mathrm{N} } \rangle 
\end{pmatrix}\,,
\label{eq:EE_BBobs}
\\
\langle C_\ell^{E_i B_j,\mathrm{o}}\rangle =& 
\vec{R}^{T}(\alpha_i, \alpha_j) \begin{pmatrix}
\langle C_\ell^{E_i E_j,\mathrm{fg}}\rangle\\
\langle C_\ell^{B_i B_j,\mathrm{fg}}\rangle 
\end{pmatrix} 
+ \vec{R}^{T}(\alpha_i+\beta, \alpha_j+\beta) \begin{pmatrix}
\langle C_\ell^{E_i E_j,\mathrm{CMB}}\rangle\\
\langle C_\ell^{B_i B_j,\mathrm{CMB}}\rangle 
\end{pmatrix}\,,
\label{eq:EBobs}
\end{align} 
where $\mathbf{R}$ and $\vec{R}$ are a rotation matrix and a rotation vector of power spectra, respectively. The explicit forms are
\begin{align}
\mathbf{R} 
( \theta_i ,  
\theta_j ) &= \begin{pmatrix}
\cos(2\theta_i) \cos(2\theta_j) & \sin(2\theta_i) \sin(2\theta_j)
\\
\sin(2\theta_i) \sin(2\theta_j) & \cos(2\theta_i) \cos(2\theta_j)
\end{pmatrix}\,,
\end{align}
and 
\begin{align}
\vec{R} (\theta_i , \theta_j)&= \begin{pmatrix}
\cos(2\theta_i)\sin(2\theta_j) 
\\
-\sin(2\theta_i) \cos(2\theta_j)
\end{pmatrix}\,.
\end{align}

Using Eqs.~(\ref{eq:EE_BBobs}) and (\ref{eq:EBobs}), 
we relate $\langle C_\ell^{E_i B_j,\mathrm{o}}\rangle$ to $\langle C_\ell^{E_i E_j,\mathrm{o}}\rangle$ and $\langle C_\ell^{B_i B_j,\mathrm{o}}\rangle$ as
\begin{align}
\begin{split}
\langle C_\ell^{E_i B_j,\mathrm{o}}\rangle &=
\vec{R}^T(\alpha_i,\alpha_j)\mathbf{R}^{-1} (\alpha_i,\alpha_j) \begin{pmatrix}
\langle C_\ell^{E_i E_j,\mathrm{o}}\rangle\\
\langle C_\ell^{B_i B_j,\mathrm{o}}\rangle 
\end{pmatrix}
-\delta_{i,j}\vec{R}^T(\alpha_i,\alpha_i)\mathbf{R}^{-1} (\alpha_i,\alpha_i) \begin{pmatrix}
\langle C_\ell^{E_i E_i,\mathrm{N}}\rangle\\
\langle C_\ell^{B_i B_i,\mathrm{N}}\rangle 
\end{pmatrix}\\
&+\left[ \vec{R}^T ( \alpha_i+\beta, \alpha_j+\beta)  -\vec{R}^T(\alpha_i,\alpha_j) \mathbf{R}^{-1}(\alpha_i,\alpha_j) \mathbf{R}(\alpha_i+\beta, \alpha_j+\beta) \right]
\begin{pmatrix}
\langle C_\ell^{E_i E_j,\mathrm{CMB}}\rangle\\
\langle C_\ell^{B_i B_j,\mathrm{CMB}}\rangle 
\end{pmatrix}
\\\Leftrightarrow
&
\begin{pmatrix}
- \vec{R}^T(\alpha_i,\alpha_j)\mathbf{R}^{-1} (\alpha_i,\alpha_j)
& 1
\end{pmatrix}
 \begin{pmatrix}
 \langle C_\ell^{E_i E_j,\mathrm{o}}\rangle\\
 \langle C_\ell^{B_i B_j,\mathrm{o}}\rangle \\
 \langle C_\ell^{E_i B_j,\mathrm{o}}\rangle
 \end{pmatrix}
 \\&-
 \left[ \vec{R}^T ( \alpha_i+\beta, \alpha_j+\beta)  -\vec{R}^T(\alpha_i,\alpha_j) \mathbf{R}^{-1}(\alpha_i,\alpha_j) \mathbf{R}(\alpha_i+\beta, \alpha_j+\beta) \right]
 \begin{pmatrix}
 \langle C_\ell^{E_i E_j,\mathrm{CMB}}\rangle\\
 \langle C_\ell^{B_i B_j,\mathrm{CMB}}\rangle 
 \end{pmatrix}
 =0\,.
\end{split}
\end{align}
In the last line we have ignored the noise term, which vanishes when $\langle C_\ell^{E_i E_i,\mathrm{N}}\rangle=\langle C_\ell^{B_i B_i,\mathrm{N}}\rangle$.

Following Eq.~(\ref{eq:LikelihoodGeneral}),
we construct a log-likelihood function as
\begin{align}\label{eq:LikelihoodCross}
\ln\mathcal{L} = -\frac{1}{2}
\sum_{\ell=2}^{\ellmax}
\left(
\mathbf{A}\vec C_\ell^{\mathrm{o}} - \mathbf{B
}\vec C_\ell^\mathrm{CMB,th}\right)^T 
\mathbf{C}^{-1}
\left(
\mathbf{A}\vec C_\ell^{\mathrm{o}} - \mathbf{B
}\vec C_\ell^\mathrm{CMB,th}\right)\,,
\end{align}
where 
$\vec C_\ell^{\mathrm{o}}$ is a one-dimensional array of $\begin{pmatrix}
 C_\ell^{E_i E_j,\mathrm{o}}&
 C_\ell^{B_i B_j,\mathrm{o}} &
 C_\ell^{E_i B_j,\mathrm{o}}
\end{pmatrix}^T$
with direct product of $i,j\in \{1,2,\cdots, N_\nu\}$,
$\vec C_\ell^\mathrm{CMB,th}$ is a one-dimensional array of 
$ \begin{pmatrix}
 C_\ell^{E_i E_j,\mathrm{CMB,th}} b_\ell^i b_\ell^j&
 C_\ell^{B_i B_j,\mathrm{CMB,th}} b_\ell^i b_\ell^j
\end{pmatrix}$,  
$\mathbf{A}$ is a block diagonal matrix of $\begin{pmatrix}
- \vec{R}^T(\alpha_i,\alpha_j)\mathbf{R}^{-1} (\alpha_i,\alpha_j)& 1
\end{pmatrix}$,
$ \mathbf{B}$ is a block diagonal matrix of $ \left[ \vec{R}^T ( \alpha_i+\beta, \alpha_j+\beta)  - \vec{R}^T(\alpha_i,\alpha_j) \mathbf{R}^{-1}(\alpha_i,\alpha_j) \mathbf{R}(\alpha_i+\beta, \alpha_j+\beta) \right]$,
and 
$\mathbf{C} = \mathbf{A}\Cov(\vec C_\ell^{\mathrm{o}},\vec C_\ell^{\mathrm{o}}{}^T)\mathbf{A}^T$.
The explicit form of $\Cov(\vec C_\ell^{\mathrm{o}},\vec C_\ell^{\mathrm{o}}{}^T)$ is 
\begin{equation}\label{eq:Cov}
\begin{split}
&\Cov(\vec{C}_\ell^{\mathrm{o}, ij}, \vec C_\ell^{\mathrm{o},pq}{}^T) \\
&=
 \begin{pmatrix}
 \Cov( C_\ell^{E_i E_j,\mathrm{o}},  C_\ell^{E_p E_q,\mathrm{o}})&\Cov( C_\ell^{E_i E_j,\mathrm{o}},  C_\ell^{B_p B_q,\mathrm{o}})&\Cov( C_\ell^{E_i E_j,\mathrm{o}}, C_\ell^{E_p B_q,\mathrm{o}})\\
 \Cov( C_\ell^{B_i B_j,\mathrm{o}}, C_\ell^{E_p E_q,\mathrm{o}})&\Cov( C_\ell^{B_i B_j,\mathrm{o}}, C_\ell^{B_p B_q,\mathrm{o}})&\Cov( C_\ell^{B_i B_j,\mathrm{o}}, C_\ell^{E_p B_q,\mathrm{o}})\\
 \Cov( C_\ell^{E_i B_j,\mathrm{o}}, C_\ell^{E_p E_q,\mathrm{o}})&\Cov( C_\ell^{E_i B_j,\mathrm{o}}, C_\ell^{B_p B_q,\mathrm{o}})&\Cov( C_\ell^{E_i B_j,\mathrm{o}}, C_\ell^{E_p B_q,\mathrm{o}})
 \end{pmatrix},
\end{split}
\end{equation}
where we use an approximate covariance for each element:
\begin{align}
\begin{split}
\Cov(C_\ell^{X,Y}, C_\ell^{Z,W}) &=
\frac{1}{(2\ell+1) }(\langle C_\ell^{X,Z} \rangle \langle C_\ell^{Y,W}\rangle + \langle C_\ell^{X,W}\rangle\langle C_\ell^{Y,Z}\rangle)   
\\&
\approx \frac{1}{(2\ell+1) }(C_\ell^{X,Z}C_\ell^{Y,W} + C_\ell^{X,W} C_\ell^{Y,Z}) .
\end{split}
\end{align}

From now on we ignore the off-diagonal elements of the covariance matrix given in Eq.~(\ref{eq:Cov}), 
because the $C_\ell^{EB,\mathrm{o}}C_\ell^{XY,\mathrm{o}}$ term with $X,Y\in\{E,B\}$  is strongly affected by the statistical fluctuations of $C_\ell^{EB,\mathrm{o}}$.
For example, let us consider $\Cov( C_\ell^{E_i B_i ,\mathrm{o}}, C_\ell^{E_i E_j,\mathrm{o}}) \sim \frac{1}{2\ell+1}(C_\ell^{E_i E_i,\mathrm{o}}C_\ell^{E_j B_i,\mathrm{o}} + C_\ell^{E_i E_j,\mathrm{o}} C_\ell^{E_i B_i,\mathrm{o}})$.
If the $j$-th frequency band is dominated by the foreground emission, $C_\ell^{E_j B_i}$ fluctuates around zero with a large statistical fluctuation due to the large $E$-mode of the foreground emission, which is enhanced by the non-zero auto frequency spectrum, $C_\ell^{E_i E_i,\mathrm{o}}$.
Thus the term $\frac{1}{(2\ell+1) }C_\ell^{E_i E_i,\mathrm{o}}C_\ell^{E_j B_i,\mathrm{o}}$ makes the approximation worse.

Even if we ignore the off-diagonal elements of Eq.~(\ref{eq:Cov}),
we find that the approximate covariance matrix sometimes differs from the sample variance especially in the low $\ell$ region. This is because we have a small number of modes in the low $\ell$ region and the approximate covariance matrix is strongly affected by the statistical fluctuation.
We reduce this fluctuation by binning power spectra in the multipole, $\ell$. Specifcally, we average the power spectra within a bin, $b$, with a bin width, $\Delta\ell$, 
and calculate the binned power spectra as
\begin{equation}\label{eq:Cb}
C_b^{X,Y} = \frac{1}{\Delta\ell}\sum_{\ell \in b} C_\ell^{X,Y}\,.
\end{equation}
The corresponding covariance matrix is given by
\begin{equation}\label{eq:CovCb}
\begin{split}
\Cov(C_b^{X,Y}, C_b^{Z,W}) &= \frac{1}{\Delta\ell^2}\sum_{\ell\in b}\Cov(C_\ell^{X,Y}, C_\ell^{Z,W})\,.
\end{split}
\end{equation}
Using these binned variables in the log-likelihood function (Eq.~\ref{eq:LikelihoodCross}),
we determine the miscalibration angles, $\alpha$, and the cosmic birefringence angle, $\beta$, simultaneously.

\section{Sky simulations}\label{sec:SkySim}
To validate our methodology,
we use the ``PySM'' package~\cite{Thorne:2016ifb} to produce realistic simulations of the microwave sky,
with experimental parameters similar to the LiteBIRD mission~\cite{Hazumi2019} (Table~\ref{tab:LBspec}).
We include polarized synchrotron and thermal dust emission with varying spectral parameters given by the ``s1'' and ``d1'' models in PySM. 
The noise is assumed to be white with standard deviation given by 
$\sigma_\mathrm{N} = (\pi/10800)(w_{\rm p}^{-1/2}/\mu{\rm K~arcmin})~\mu \mathrm{K~str^{-1/2}} $~\cite{Katayama:2011eh}
with $w_{\rm p}^{-1/2}$ given in the ``Polarization sensitivity'' column of Table~\ref{tab:LBspec}.
A CMB map is generated from the power spectra calculated by CAMB~\cite{Lewis:2000}
using the latest Planck 2018 cosmological parameters for ``TT,TE,EE$+$lowE$+$lensing''~\cite{Aghanim:2018eyx}:
$\Omega_bh^2=0.02237$, $\Omega_ch^2=0.1200$,
$h=0.6736$, $\tau=0.0544$, $A_s=2.100\times 10^{-9}$, and $n_s=0.9649$.
The CMB map includes the lensed $B$-mode but does not include the primordial $B$-mode,
i.e., the tensor-to-scalar ratio, $r$, is zero.

We use the HEALPix package~\cite{Gorski:2004by} to deal with maps, spherical harmonics coefficients, and power spectra.
We incorporate the beam smearing of the LiteBIRD telescope 
by multiplying the spherical harmonics coefficients of the CMB and foreground maps at each frequency 
by  the appropriate beam transfer function, $b_\ell$. Specifically, we use a Gaussian beam with full-width-half-maximum (FWHM) given in the third column of Table~\ref{tab:LBspec}.
We use the map resolution parameter of $N_\mathrm{side} =512$ and
 calculate the power spectra from $\ell_\mathrm{min}=2$ to $\ell_\mathrm{max}= 2 N_\mathrm{side} = 1024$.

\begin{table}
	\centering
	\caption{Polarization sensitivity and beam size of the LiteBIRD telescopes~\cite{Hazumi2019}}\label{tab:LBspec}
	\begin{tabular}{c c c}
		\toprule
		Frequency (GHz) & Polarization sensitivity ($\mathrm{\mu K^{'}}$) & Beam size in FWHM (arcmin) \\ 
		\midrule
		40 & 37.5 & 69 \\
		50 & 24.0 & 56 \\
		60 & 19.9 & 48 \\
		68 & 16.2 & 43 \\
		78 & 13.5 & 39 \\
		89 & 11.7 & 35 \\
		100 & 9.2 & 29 \\
		119 & 7.6 & 25 \\
		140 & 5.9 & 23 \\
		166 & 6.5 & 21 \\
		195 & 5.8 & 20 \\
		235 & 7.7 & 19 \\
		280 & 13.2 & 24 \\
		337 & 19.5 & 20\\
		402 & 37.5 & 17 \\
		\bottomrule
	\end{tabular}
\end{table}

We generate the input miscalibration angles, $\alpha$, 
randomly from a Gaussian distribution with zero mean and standard deviation of $0.33\,\deg$ ($19.8\,$arcmin),
which corresponds to the calibration uncertainty of the Crab Nebula (Tau A)~\cite{Aumont:2018epb}.

In the following sections,
we show the results from both a single realization and Monte Carlo (MC) realizations. For the single realization, we use the actual distribution of the Galactic foreground given by PySM. For MC realizations, we use the \texttt{synfast} function of HEALPix to generate Gaussian random realizations of the Galactic foreground emission from the power spectra of the PySM maps. Specifically, we generate map realizations at the reference frequency bands ($353\,$GHz for the thermal dust emission and $23\,$GHz for the synchrotron emission), and scale these maps to each frequency band with the same model used in PySM. If we did not do this, all of the MC realizations would have the same foreground sky, which would underestimate the sample variance. The goal of the MC realizations is to check whether the estimated values of $\alpha$ and $\beta$ are unbiased, and the estimated values of their uncertainties are also unbiased.
\section{Results}\label{sec:Results}
\subsection{\label{sec:alpha}Determination of \texorpdfstring{$\alpha$}{alpha}}
\begin{table}
\centering
\caption{Input ($\alpha_\mathrm{in}$) and recovered ($\alpha_\mathrm{out}$) miscalibration angles and their 68\% uncertainties ($\sigma(\alpha_\mathrm{out})$).
	We show the results from the cross frequency correlation (Eq.~(\ref{eq:LikelihoodCross})) and the auto correlation power spectra (Eq.~(\ref{eq:LikelihoodGeneral})). For the MC simulations we show the mean of $\alpha_\mathrm{out}$ and its standard deviation, as well as the mean of $\sigma(\alpha_\mathrm{out})$ estimated from  Eq.~(\ref{eq:LikelihoodCross}) for each realization.
}
\label{tab:LBResults}
\begin{tabular}{ccccccccc}
%\begin{tabular}{rrrrrrrr}
\toprule
\multirow{3}{*}{$\nu$ (GHz)} & \multirow{3}{*}{\shortstack[c]{$\alpha_\mathrm{in}$ \\ (arcmin)} } & \multicolumn{5}{c}{\shortstack[c]{$\alpha_\mathrm{out}$  (arcmin) \\ with cross correlation (Eq.~\ref{eq:LikelihoodCross}) }} & \multicolumn{2}{c}{\shortstack[c]{$\alpha_\mathrm{out}$ (arcmin) \\ with auto correlation (Eq.~\ref{eq:LikelihoodGeneral}) } }  \\
 \cmidrule(lr){3-7} \cmidrule(lr){8-9}
&& \multicolumn{2}{c}{Single realization} & \multicolumn{3}{c}{MC simulations}&\multicolumn{2}{c}{Single realization} \\
 \cmidrule(lr){3-4} \cmidrule(lr){5-7} \cmidrule(lr){8-9}
%\midrule
&&$\alpha_\mathrm{out}$ & $\sigma(\alpha_\mathrm{out})$ & $\bar{\alpha}_\mathrm{out}$ & $\mathrm{std}(\alpha_\mathrm{out})$ & $\bar{\sigma}(\alpha_\mathrm{out})$ &$\alpha_\mathrm{out}$ & $\sigma(\alpha_\mathrm{out})$\\
\midrule
  40 &    7.2 &    9.0 &    2.7 &    7.5 &   2.4 &  2.5 &   11.7 &  8.5 \\
  50 &    4.2 &    3.7 &    2.2 &    4.6 &   1.9 &  2.0 &    1.7 &  6.3 \\
  60 &    1.8 &    0.5 &    1.7 &    2.1 &   1.8 &  1.7 &    4.1 &  4.9 \\
  68 &  -29.4 &  -26.0 &    1.5 &  -28.8 &   1.3 &  1.5 &  -30.6 &  3.5 \\
  78 &   24.0 &   25.2 &    1.1 &   24.1 &   1.1 &  1.0 &   28.5 &  2.6 \\
  89 &  -16.8 & -16.05 &   0.75 & -16.52 &  0.77 & 0.72 &  -16.1 &  1.9 \\
 100 &   24.0 &  24.21 &   0.51 &  24.03 &  0.54 & 0.51 &   25.5 &  1.2 \\
 119 &   -3.6 &  -2.62 &   0.38 &  -3.52 &  0.38 & 0.37 &   -2.9 &  1.0 \\
 140 &   15.6 &  15.17 &   0.32 &  15.56 &  0.32 & 0.32 &   15.7 &  1.0 \\
 166 &   16.2 &  16.28 &   0.32 &  16.15 &  0.31 & 0.32 &   17.6 &  1.2 \\
 195 &  -36.6 & -36.55 &   0.34 & -36.57 &  0.30 & 0.34 &  -37.6 &  1.5 \\
 235 &   32.4 &  32.74 &   0.37 &  32.29 &  0.34 & 0.37 &   30.5 &  2.4 \\
 280 &  -10.2 &  -9.86 &   0.39 & -10.24 &  0.34 & 0.38 &  -12.2 &  4.6 \\
 337 &   49.2 &  49.67 &   0.45 &  49.03 &  0.38 & 0.45 &   46.4 &  4.8 \\
 402 &  -40.2 & -39.73 &   0.45 & -40.25 &  0.37 & 0.45 &  -47.8 &  4.5 \\
\bottomrule
\end{tabular}
\end{table}

First we report the results of the determination of $\alpha$ in the absence of $\beta$, using the log-likelihood given in Eq.~(\ref{eq:LikelihoodCross}). We use the \texttt{anafast} function of HEALPix to calculate cross frequency spectra over the full sky. We use the bin size of $\Delta \ell = 20$,
and calculate the binned power spectra and the covariance matrices using Eqs.~(\ref{eq:Cb}) and (\ref{eq:CovCb}), respectively.

In Table~\ref{tab:LBResults},
we show the input ($\alpha_\mathrm{in}$) and recovered ($\alpha_\mathrm{out}$) miscalibration angles, the uncertainty derived from a single realization using the likelihood ($\sigma(\alpha_\mathrm{out})$), the mean of $\alpha_\mathrm{out}$ estimated from the MC realizations ($\bar\alpha_\mathrm{out}$), its standard deviation (std($\alpha_\mathrm{out}$)), and the mean of the uncertainties derived from the MC realizations using the likelihood ($\bar\sigma(\alpha_\mathrm{out})$).
Here, $\alpha_\mathrm{out}$ and $\sigma(\alpha_\mathrm{out})$ are the median and the  $1\sigma~(68\%)$ volume of the marginalized distribution, respectively.
We find that the recovered angles are unbiased.
We also find that $\sigma(\alpha_\mathrm{out})$ agrees with std($\alpha_\mathrm{out})$, although the former is slightly overestimated in the highest frequency bands.

The last two columns in Table~\ref{tab:LBResults} show the results using only the auto frequency spectra with the likelihood given in Eq.~(\ref{eq:LikelihoodGeneral}) \cite{Minami:2019ruj}. Comparing to them, we find that the cross spectra reduce  $\sigma(\alpha_\mathrm{out})$ dramatically, by at least a factor of two, up to an order of magnitude. The benefit of including the cross spectra in the analysis is obvious. The signal-to-noise of the miscalibration angle at a given band can be most efficiently improved by cross-correlating it with the other bands with lower noise. Therefore the improvement is most dramatic in the lowest and highest frequency bands. 

\begin{figure}
	\centering
	\includegraphics[width=\linewidth]{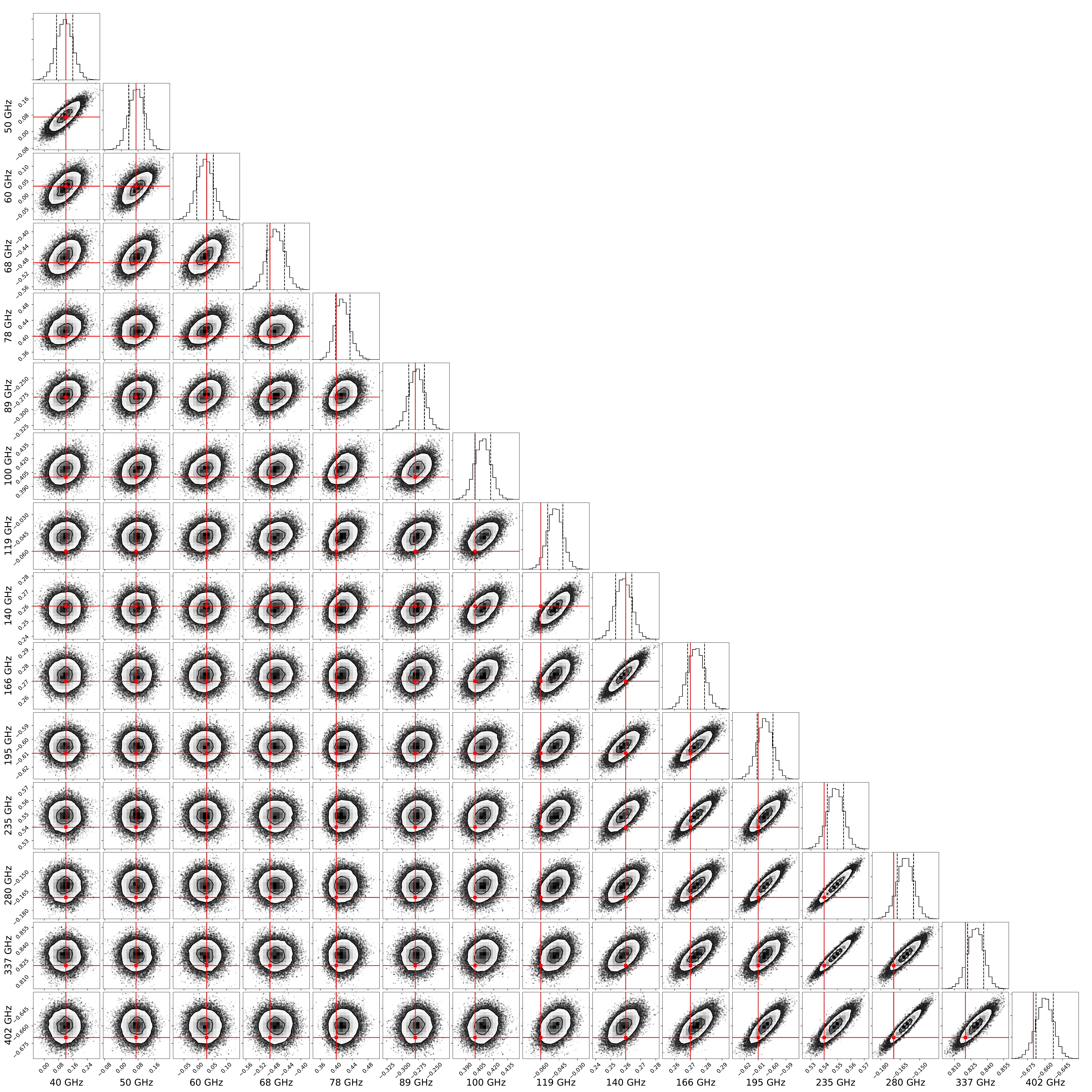}
	\caption{Posterior distributions of the recovered miscalibration angles, $\alpha$, from the single realization with the experimental parameters given in Table~\ref{tab:LBspec}.
	The solid lines in the 2D contours show the 1$\sigma$ ($39.3\%$) and 2$\sigma$ ($86.5\%$) quantiles.
	The dashed lines in the 1D histograms show the $1\sigma$ (from $16\%$ to $84\%$) quantiles, while the red lines show the input miscalibration angles, $\alpha_\mathrm{in}$,
	 given in Table~\ref{tab:LBResults}.
	}
	\label{fig:MCMC_alpha}
\end{figure}

The recovered angles $\alpha$ at different frequency bands are correlated, since the CMB and foregrounds components are common. To see the correlation,
we show the posterior distributions of the miscalibration angles of all frequency bands in Figure~\ref{fig:MCMC_alpha}.
The angles in the foreground-dominated bands (i.e., lowest and highest frequency bands) are most tightly correlated, as the uncertainties are dominated by the same foreground (synchrotron and thermal dust emission in the lowest and highest frequency bands, respectively). In the mid frequency bands (e.g., around 119\,GHz), the angles in the adjacent frequencies are correlated because they have a similar signal-to-noise for the CMB. The correlation weakens for widely-separated frequency combinations. 

%\section{Simultaneous determination of cosmic birefringence and miscalibration angles}
\subsection{Simultaneous determination of \texorpdfstring{$\alpha$ and $\beta$}{alpha and beta}}
\label{sec:Birefringence}
Next, we report the results of simultaneous determination of $\alpha$ and $\beta$ using the log-likelihood given in Eq.~(\ref{eq:LikelihoodCross}).
The input value of the cosmic birefringence angle is $\beta = 30$\,arcmin, which is the current upper bound~\cite{Aghanim:2016fhp}.
We use the same input miscalibration angles as in Sect.~\ref{sec:alpha} and use 100 MC samples as we described in Sect.~\ref{sec:SkySim}.

We show the input and recovered values of $\beta$ and $\alpha$ as well as their uncertainties in Table~\ref{tab:AlphaBeta}. We find that the uncertainty in $\beta$ from the cross frequency power spectra is more than a factor of two smaller than 11\,arcmin obtained from the auto frequency spectra \cite{Minami:2019ruj}. Therefore, an experiment such as LiteBIRD can significantly improve upon the constraint on $\beta$ compared to the current limit.

In detail, when we use all the 15 frequency bands, we find that $\beta$ is biased slightly low, and the uncertainty of $\beta$ derived from the log-likelihood given in Eq.~(\ref{eq:LikelihoodCross}) is slightly smaller than the standard deviation of the MC realizations. 
While we need both the CMB and the Galactic foreground emission to distinguish between $\alpha$ and $\beta$, including the bands that are completely dominated by the foreground appears to bias the results. We mitigate this by removing the foreground-dominated frequency bands.
For example, we show the results from six CMB-dominated bands, from $89\,$GHz to $196\,$GHz in the right three columns of  Table~\ref{tab:AlphaBeta}. 
It shows that we can correctly recover  $\alpha$ and $\beta$ simultaneously with consistent uncertainties. Even in this conservative case the uncertainty of $\beta$ is 6\,arcmin, which is nearly a factor of two smaller than the auto frequency spectra.

\begin{figure}
	\centering
	\includegraphics[width=\linewidth]{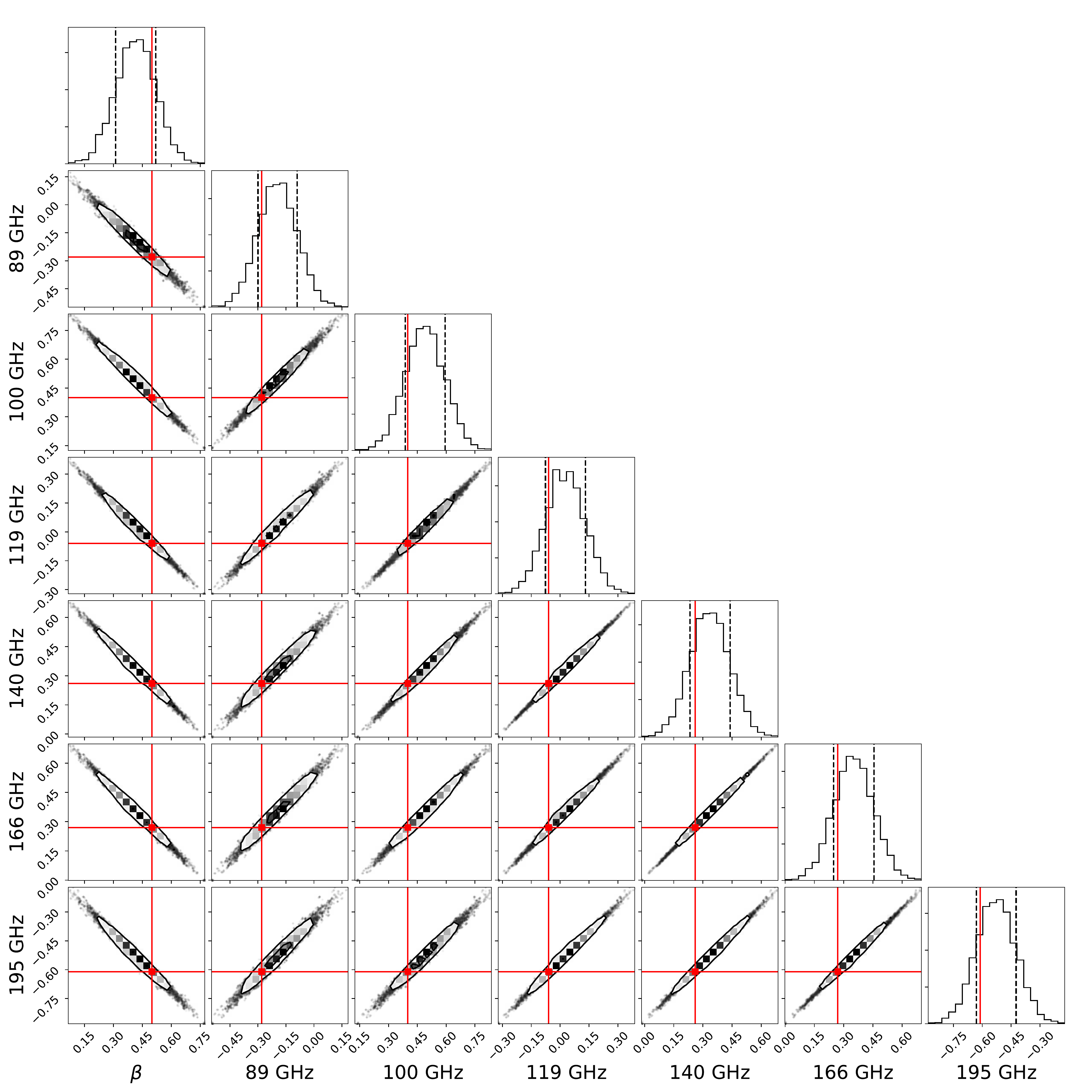}
	\caption{Similar to Figure~\ref{fig:MCMC_alpha} but for the recovered cosmic birefringence ($\beta$) and miscalibration angles at the indicated frequency bands
	from the single realization at six CMB-dominated bands.
}
\label{fig:MCMC_alphabeta}
\end{figure}

Our method uses the CMB signal to determine $\alpha+\beta$ and lifts the degeneracy by the foreground signal which only determines $\alpha$. Therefore, there are linear correlations between $\alpha$ and $\beta$ (Figure~\ref{fig:MCMC_alphabeta}), and the uncertainties of $\alpha$ and $\beta$ are similar (Table~\ref{tab:AlphaBeta}), as pointed out by Ref.~\cite{Minami:2019ruj}.

\begin{table}
\centering
\caption{
	Input and recovered cosmic birefringence and miscalibration angles, $\theta=(\beta,\alpha)$, and their 68\% uncertainties from the cross frequency power spectra (Eq.~(\ref{eq:LikelihoodCross})) with 100 MC samples.
	We show the mean of $\theta_\mathrm{out}$ and its standard deviation, as well as the mean of $\sigma(\theta_\mathrm{out})$ estimated from  Eq.~(\ref{eq:LikelihoodCross}) for each realization.
	We show two results from
	(1) all 15 frequency bands, and
	(2) six CMB-dominated bands.
}
\label{tab:AlphaBeta}
\begin{tabular}{cccccccc}
\toprule
\multirow{2}{*}{ \shortstack{$\nu$ (GHz)\\ or $\beta$}} & \multirow{2}{*}{\shortstack[c]{$\theta_\mathrm{in}$ \\ (arcmin)} } &
 \multicolumn{3}{c}{\shortstack[c]{ Recovered angles (arcmin) \\ with all bands }} & \multicolumn{3}{c}{\shortstack[c]{Recovered angles (arcmin)\\ with CMB-dominated bands}}\\
 \cmidrule(lr){3-5} \cmidrule(lr){6-8}
&& $\bar{\theta}_\mathrm{out}$& $\mathrm{std} ( \theta_\mathrm{out} )$ & $ \bar{\sigma }_\mathrm{out}$ & $\bar{\theta}_\mathrm{out}$& $ \mathrm{std} (\theta_\mathrm{out})$ & $\bar{\sigma}_\mathrm{out}$\\
\midrule
$\beta$ &  30.0 &  24.9 & 4.7  & 3.6 &  28.9 & 6.3 & 6.1 \\
 40     &   7.2 &  12.2 & 5.1  & 4.2 &     - &   - &   - \\
 50     &   4.2 &   9.3 & 4.9  & 4.0 &     - &   - &   - \\
 60     &   1.8 &   7.0 & 4.9  & 3.9 &     - &   - &   - \\
 68     & -29.4 & -24.0 & 4.9  & 3.9 &     - &   - &   - \\
 78     &  24.0 &  29.0 & 4.7  & 3.7 &     - &   - &   - \\
 89     & -16.8 & -11.5 & 4.7  & 3.7 & -15.5 & 6.5 & 6.1 \\
100     &  24.0 &  28.9 & 4.7  & 3.6 &  25.0 & 6.3 & 6.0 \\
119     &  -3.6 &   1.4 & 4.7  & 3.6 &  -2.5 & 6.3 & 6.0 \\
140     &  15.6 &  20.5 & 4.7  & 3.6 &  16.5 & 6.2 & 6.0 \\
166     &  16.2 &  21.1 & 4.7  & 3.6 &  17.0 & 6.2 & 6.0 \\
195     & -36.6 & -31.6 & 4.7  & 3.6 & -35.6 & 6.2 & 6.0 \\
235     &  32.4 &  37.2 & 4.7  & 3.6 &     - &   - &   - \\
280     & -10.2 &  -5.2 & 4.7  & 3.6 &     - &   - &   - \\
337     &  49.2 &  53.9 & 4.7  & 3.6 &     - &   - &   - \\
402     & -40.2 & -35.2 & 4.7  & 3.6 &     - &   - &   - \\
\bottomrule
\end{tabular}
\end{table}

\section{Discussion and conclusion}\label{sec:Conclusion}
In this paper, we have developed a strategy to determine the cosmic birefringence ($\beta$) and miscalibrated polarization angles ($\alpha$) of CMB experiments using the observed $EB$ power spectra of the CMB and Galactic foreground emission. We have extended the methodology of Ref.~\cite{Minami:2019ruj}, which was developed for the auto frequency power spectra, by including the cross frequency spectra. The idea is simple:
when we use the cross correlation between detectors with small and large noise,
we can effectively reduce the calibration uncertainty on the detector with large noise.

Applying our method to simulated maps of CMB, 
realistic foreground emission~\cite{Thorne:2016ifb}, and instrumental noise
with beam smearing similar to the LiteBIRD~\cite{Hazumi2019},
we have found that 
the method correctly recovers the input values of $\alpha$ and $\beta$, with significantly smaller uncertainties on both
$\alpha$ and $\beta$ compared to the auto frequency spectra. If we ignore $\beta$, the uncertainties on $\alpha$ are reduced by at least a factor of two, up to an order of magnitude depending on the noise levels of frequency bands. If we include $\beta$, the uncertainties on $\beta$ and $\alpha$ become similar as the CMB tightly constrains $\alpha+\beta$ while the foreground constrains $\alpha$, as pointed out by Ref.~\cite{Minami:2019ruj}. We find that the cross frequency spectra reduce the uncertainty of $\beta$ by a factor of two, yielding $\sigma(\beta)\simeq 6$\,arcmin; thus, an experiment such as LiteBIRD can significantly improve upon the current limit on the cosmic birefringence, 30\,arcmin \cite{Aghanim:2016fhp}.

While we have applied our method to the LiteBIRD specification,
we can apply the method to any multi frequency observations. The extension to ground-based experiments with a partial sky coverage can be done following Ref.~\cite{Minami:2020xfg}. This new framework allows us to reduce the uncertainty in the miscalibration polarization angle and to detect the cosmic birefringence.

%\section{Enunciations}
%%%% Most of the enunciations like theorem, lemma, corollary, proposition, defintion,
%%%% condition, example, conjecture etc. are defined in the class file.

%%%% If the author wants to add or modify the enunciation style
%%%% they can define in the preamble as shown below.

%%%% \newtheoremstyle{theorem}{6pt}{6pt}{\rm}{}{\sffamily}{ }{ }{}
%%%% \theoremstyle{theorem}
%%%% \newtheorem{theorem}{\sc Theorem}[section]

%%%%\newtheoremstyle{corollary}{6pt}{6pt}{\rm}{}{\sffamily}{ }{ }{}
%%%%\theoremstyle{corollary}
%%%%\newtheorem{corollary}{\sc Corollary}[section]

%%%%\newtheoremstyle{definition}{6pt}{6pt}{\rm}{}{\sffamily}{ }{ }{}
%%%%\theoremstyle{definition}
%%%%\newtheorem{definition}[theorem]{\sc Definition}
%%%%
%%%%\newtheorem{exercise}[theorem]{Exercise}

\section*{Acknowledgment}
We thank Y. Chinone, K. Ichiki, N. Katayama, T. Matsumura, H. Ochi, S. Takakura, and Joint Study Group of the LiteBIRD collaboration for useful discussions. This work was supported in part by the Japan Society for the Promotion of Science (JSPS) KAKENHI, Grant Number JP20K1449 and JP15H05896, and the Excellence Cluster ORIGINS which is funded by the Deutsche Forschungsgemeinschaft (DFG, German Research Foundation) under Germany’s Excellence Strategy
- EXC-2094 - 390783311. The Kavli IPMU is supported by World Premier International Research Center Initiative (WPI), MEXT, Japan.
\bibliographystyle{ptephy}
\bibliography{references}

\begin{thebibliography}{10}

\bibitem{Carroll:1998zi}
S.~M. Carroll, Phys. Rev. Lett., {\bf 81}, 3067--3070 (1998),
  {{arXiv:astro-ph/9806099}}.

\bibitem{Lue:1998mq}
A.~Lue, L.-M. Wang, and M.~Kamionkowski, Phys. Rev. Lett., {\bf 83}, 1506--1509
  (1999),  {{arXiv:astro-ph/9812088}}.

\bibitem{Feng:2004mq}
B.~Feng, H.~Li, M.~Li, and X.~Zhang, Phys. Lett., {\bf B620}, 27--32 (2005),
  {{arXiv:hep-ph/0406269}}.

\bibitem{Feng:2006dp}
B.~Feng, M.~Li, J.-Q. Xia, X.~Chen, and X.~Zhang, Phys. Rev. Lett., {\bf 96},
  221302 (2006),  {{arXiv:astro-ph/0601095}}.

\bibitem{Liu:2006uh}
G.-C. Liu, S.~Lee, and K.-W. Ng, Phys. Rev. Lett., {\bf 97}, 161303 (2006),
  {{arXiv:astro-ph/0606248}}.

\bibitem{Saito:2007kt}
S.~Saito, K.~Ichiki, and A.~Taruya, JCAP, {\bf 0709}, 002 (2007),
  {{arXiv:0705.3701}}.

\bibitem{Keating:2012ge}
B.~Keating, M.~Shimon, and A.~Yadav, Astrophys. J., {\bf 762}, L23 (2012),
  {{arXiv:1211.5734}}.

\bibitem{Komatsu:2010fb}
E.~Komatsu et~al., Astrophys. J. Suppl., {\bf 192}, 18 (2011),
  {{arXiv:1001.4538}}.

\bibitem{Minami:2019ruj}
Yuto Minami, Hiroki Ochi, Kiyotomo Ichiki, Nobuhiko Katayama, Eiichiro Komatsu,
  and Tomotake Matsumura, PTEP, {\bf 2019}(8), 083E02 (2019),
  {{arXiv:1904.12440}}.

\bibitem{Hazumi2019}
M.~Hazumi et~al., Journal of Low Temperature Physics, {\bf 194}(5), 443--452
  (Mar 2019).

\bibitem{PhysRevD.80.043522}
L.~Pagano, P.~de~Bernardis, G.~De~Troia, G.~Gubitosi, S.~Masi, A.~Melchiorri,
  P.~Natoli, F.~Piacentini, and G.~Polenta, Phys. Rev., {\bf D80}, 043522
  (2009),  {{arXiv:0905.1651}}.

\bibitem{Gubitosi:2014cua}
G.~Gubitosi, M.~Martinelli, and L.~Pagano, JCAP, {\bf 1412}(12), 020 (2014),
  {{arXiv:1410.1799}}.

\bibitem{MOLINARI201665}
A.~Gruppuso D.~Molinari and P.~Natoli, Physics of the Dark Universe, {\bf 14},
  65 -- 72 (2016).

\bibitem{Gruppuso_2016}
A.~Gruppuso, M.~Gerbino, P.~Natoli, L.~Pagano, N.~Mandolesi, A.~Melchiorri, and
  D.~Molinari, JCAP, {\bf 2016}(06), 001--001 (jun 2016).

\bibitem{Gruppuso:2016nhj}
A.~Gruppuso, G.~Maggio, D.~Molinari, and P.~Natoli, JCAP, {\bf 2016}(05),
  020--020 (may 2016).

\bibitem{Pogosian:2019jbt}
Levon Pogosian, Meir Shimon, Matthew Mewes, and Brian Keating, Phys. Rev. D,
  {\bf 100}(2), 023507 (2019),  {{arXiv:1904.07855}}.

\bibitem{Minami:2020xfg}
Yuto Minami, PTEP, {\bf 2020}(6), 063E01 (2020),  {{arXiv:2002.03572}}.

\bibitem{planckdust:2016}
Planck Collaboration~Int. XXX, Astron. Astrophys., {\bf 586}, A133 (2016),
  {{arXiv:1409.5738}}.

\bibitem{planckdust:2018}
Planck~Collaboration XI, Astron. Astrophys. (2018),  {{arXiv:1801.04945}}.

\bibitem{Aghanim:2016fhp}
Planck Collaboration~Int. XLIX, Astron. Astrophys., {\bf 596}, A110 (2016),
  {{arXiv:1605.08633}}.

\bibitem{Abitbol:2015epq}
M.~H. Abitbol, J.~C. Hill, and B.~R. Johnson, Mon. Not. Roy. Astron. Soc., {\bf
  457}(2), 1796--1803 (2016),  {{arXiv:1512.06834}}.

\bibitem{Thorne:2017jft}
B.~Thorne, T.~Fujita, M.~Hazumi, N.~Katayama, E.~Komatsu, and M.~Shiraishi,
  Phys. Rev., {\bf D97}(4), 043506 (2018),  {{arXiv:1707.03240}}.

\bibitem{Thorne:2016ifb}
B.~Thorne, J.~Dunkley, D.~Alonso, and S.~Naess, Mon. Not. Roy. Astron. Soc.,
  {\bf 469}(3), 2821--2833 (2017),  {{arXiv:1608.02841}}.

\bibitem{Katayama:2011eh}
N.~Katayama and E.~Komatsu, Astrophys. J., {\bf 737}, 78 (2011),
  {{arXiv:1101.5210}}.

\bibitem{Lewis:2000}
A.~{Lewis}, A.~{Challinor}, and A.~{Lasenby}, Astrophys. J., {\bf 538},
  473--476 (August 2000),  {{astro-ph/9911177}}.

\bibitem{Aghanim:2018eyx}
N.~Aghanim et~al., Astron. Astrophys. (2018),  {{arXiv:1807.06209}}.

\bibitem{Gorski:2004by}
K.~M. Gorski, E.~Hivon, A.~J. Banday, B.~D. Wandelt, F.~K. Hansen, M.~Reinecke,
  and M.~Bartelman, Astrophys. J., {\bf 622}, 759--771 (2005),
  {{arXiv:astro-ph/0409513}}.

\bibitem{Aumont:2018epb}
Jonathan Aumont, Juan~Francisco Macías-Pérez, Alessia Ritacco, Nicolas
  Ponthieu, and Anna Mangilli, Astron. Astrophys., {\bf 634}, A100 (2020),
  {{arXiv:1805.10475}}.

\end{thebibliography}
%
% once the .bbl file has been generated then place the text in your article.
\appendix
\end{document}